\documentclass{article}

\begin{document}

\title{Multiple Photonic Shells Around a Line Singularity}
\author{M. Arik\footnote{Bogazici University, Department of Physics, Bebek, Istanbul, Turkey; e-mail:arikm@boun.edu.tr}   and O. Delice\footnote{Bogazici University, Department of Physics, Bebek, Istanbul, Turkey; e-mail:odelice@boun.edu.tr}}
\maketitle
\begin{abstract}
Line singularities  including cosmic strings may be screened by
photonic shells until they appear as a planar wall.
\end{abstract}
\\

Infinitely thin static photonic shells composed of equal amounts
of oppositely moving photons following circular, helical or axial
trajectories around a cosmic string \cite{bicak} and, around a
general line singularity \cite{ArikDelice} have recently been
found. In this paper, we will find solutions giving multiple
photonic shells around a line singularity. Multiple photonic
shells around cosmic strings will be obtained as a limiting case.
We will show that the exterior region of the (multiple) shells may
become an infinite plane wall.

The vacuum spacetime exterior to a static, cylindrically symmetric
source is described by the Levi-Civita metric \cite{levicivita}
which can be written as:

\begin{equation}
ds^{2}=-\rho ^{4\sigma }dt^{2}+\rho ^{4\sigma (2\sigma -1)}(d\rho
^{2}+ P^2dz^{2})+Q^2\rho ^{2(1-2\sigma )}d\phi^{2} \label{Lc}
\end{equation}
where $t$ , $\rho $, $z$ and $\phi$ are the time, the radial, the
axial and the angular variables of the cylindrical coordinates
with ranges $-\infty <t,z< \infty ,$ $0 \leq \rho < \infty$ and
$0\leq\phi \leq 2\pi$.  $\sigma$, $P$ and $Q$ are real constants.
Transforming the radius $\rho $ into a proper radius $r$ by
defining
$
dr=\rho ^{2\sigma (2\sigma -1)}d\rho \,
$ 
puts 
(\ref{Lc}) in the Kasner form:

\begin{equation}
ds^{2}=-(\frac{r}{r_{0}})^{2a}dt^{2}+dr^{2}+(\frac{r}{r_{0}})^{2b}dz^{2}+(%
\frac{r}{r_{0}})^{2c}r_{0}^{2}\alpha ^{2}d\phi ^{2},
\label{kasner}
\end{equation}
where
\begin{equation}
\rho =(Nr)^{1/N},
\  N=4\sigma ^{2}-2\sigma +1,\
  a=\frac{2\sigma}{N},\  \
b=\frac{-2\sigma (1-2\sigma )}{N},\  c=\frac{1-2\sigma
}{N}. \  
\end{equation}
Here $\alpha ,a,b,c$ are real constants and $r$ is the radial
coordinate. Note that we rescaled the $z$ and $t$ coordinates and
the constant $\alpha $ is given by:
\begin{equation}\label{alpha}
\alpha=QN^c(r_o)^{c-1}.
\end{equation}

 This metric is  cylindrically symmetric and
represents the exterior field of the line singularity at $r=0$
\cite{israel}. The Einstein tensor for this metric gives vacuum
solutions of Einstein equations with the constraints
\begin{equation}a+b+c=a^{2}+b^{2}+c^{2}=1,
\label{abcrel}
\end{equation}
so only one of the parameters $a,b,c$ is free. In this paper we
need $a,c>0$ and $b<0$ to have a source with positive energy. For
detailed discussion of the properties of the metrics (\ref{Lc})
and (\ref{kasner}) see \cite{ArikDelice} and references therein.
The metric reduces to the famous cosmic string metric
\cite{vilenkin} when $a=b=0, c=1$ keeping  the angular defect
parameter $\alpha < 1$.

Let us first review the photonic shell solution around a line
singularity. We choose the interior and the exterior regions of
the infinitely long thin cylindrical shell with radius $r_1$ to be
two different Levi-Civita metrics in Kasner form with the metrics:
\begin{equation}
ds_{-}^{2}=-(\frac{r}{r_{1}})^{2a}dt^{2}+dr^{2}+(\frac{r}{r_{1}}
)^{2b}dz^{2}+(\frac{r}{r_{1}})^{2c}\alpha ^{2}r_{1}^{2} d\phi
^{2}\  \ \ (r<r_{1}),\label{interiorkmetric}
\end{equation}
and
\begin{equation}
ds_{+}^{2}=-(\frac{r}{r_{1}})^{2a'}dt^{2}+dr^{2}+(\frac{r}{r_{1}}
)^{2b'}dz^{2} +(\frac{r}{r_{1}})^{2c'}\alpha ^{2}r_{1}^{2} d\phi
^{2} \ \ \ (r>r_{1}), \label{exteriorkmetric}
\end{equation}
where $a,b,c$ and $a',b',c'$ satisfy the relations (\ref{abcrel}).
We can define an infinitely thin and long shell if the metrics are
continuous everywhere but their first derivatives are
discontinuous at $r=r_1$. These discontunitues may give rise to an
infinitely thin shell.
 We can combine
(\ref{interiorkmetric}) and (\ref{exteriorkmetric}) in the form :

\begin{equation}
ds^{2}=-A^{2}(r)dt^{2}+dr^{2}+B^{2}(r)dz^{2}+C^{2}(r)d\phi ^{2}
\label{metricABC}
\end{equation}
with
\begin{equation}\label{A1}
A(r)=(r/r_{1})^{a}\ \theta (r_{1}-r)+(r/r_{1})^{a'}\ \theta
(r-r_{1}),
\end{equation}

\begin{equation}
B(r)=(r/r_{1})^{b}\ \theta
(r_{1}-r)+(r/r_{1})^{b'}\ \theta (r-r_{1}),
\end{equation}
 and
\begin{equation}
C(r)=[(r/r_{1})^{c}\ \theta (r_{1}-r)+(r/r_{1})^{c'}\ \theta (r-r_{1})]r_{1}\alpha ,
\end{equation}
where $\theta (x-x_{0})$ is the Heaviside step function.

 The nonzero components of the Einstein tensor for the metric (\ref{metricABC})
are given by

\begin{equation}G_{00}=-(\frac{B_{rr}}{B}+\frac{C_{rr}}{C}+\frac{B_{r}C_{r}}{BC})
\end{equation}

\begin{equation}G_{rr}=G_{11}=\frac{A_{r}B_{r}}{AB}+\frac{A_{r}C_{r}}{AC}+\frac{B_{r}C_{r}}{
BC}
\end{equation}

\begin{equation}G_{zz}=G_{22}=\frac{A_{rr}}{A}+\frac{C_{rr}}{C}+\frac{A_{r}C_{r}}{AC}
\end{equation}

\begin{equation}G_{\phi \phi }=G_{33}=\frac{A_{rr}}{A}+\frac{B_{rr}}{B}+\frac{A_{r}B_{r}}{AB
},
\end{equation}
where subscripts denote partial derivatives. Since the exterior
and interior regions are vacuum, the only surviving terms
are the terms which contain Dirac delta functions giving 
the energy momentum tensor of the shell. The nonzero elements of
$G_{\mu \nu }$ are

\begin{equation}G_{00}=-\frac{b'-b+c'-c}{r_{1}}\ \delta (r-r_{1})=\frac{%
a'-a}{r_{1}}\ \delta (r-r_{1}),
\end{equation}

\begin{equation}G_{22}=\frac{a'-a+c'-c}{r_{1}}\ \delta (r-r_{1})=\frac{%
b-b'}{r_{1}}\ \delta (r-r_{1}), \label{G22shell}
\end{equation}

\begin{equation}G_{33}=\frac{a'-a+b'-b}{r_{1}}\ \delta (r-r_{1})=\frac{%
c-c'}{r_{1}}\ \delta (r-r_{1})\label{G33shell}.
\end{equation}

For the Einstein equation
\begin{equation}
G_{\mu\nu}=8 \pi G \ T_{\mu\nu}
\end{equation}
we can choose the energy momentum tensor of the shell in the form
\begin{equation}
T_{\mu\nu}=diag(\rho,p_r,p_z,p_\phi)\label{enmomtensor}
\end{equation}
where $\rho$ is the energy density  and $p_i \ (i=1,2,3)$ are the
principal pressures. 
Since we used an orthonormal basis, we have $T^0_0=-T_{00}$ and
$T^i_j=T_{ij}\ (i,j=r,z,\phi) $. Using (\ref{abcrel}) one can show
that the energy momentum tensor of the shell satisfies the
condition

\begin{equation}\label{traceless}
 T^\mu_\mu =0
\end{equation}
 and this result can
be interpreted as an infinitely long thin shell along the $z$
direction with radius $r_1$ composed of equal amount of oppositely
moving photons along a helical direction around a line
singularity. This helical motion gives rise to pressures in the
$\phi$ and $z$ directions with the equation of state
$\rho=p_z+p_\phi.$ Thus if one chooses the interior and the
exterior metrics of the shell as Levi-Civita metrics in Kasner
form (\ref{interiorkmetric},\ref{exteriorkmetric}) then the shell
is necessarily composed of massless particles. Since we have
$a,a'>0$ if $a'>a$ the shell has positive energy density. The
singularity at $r=0$ has positive effective mass density. Choosing
$b=b'$ in (\ref{G22shell} ) give rise to the solution
$\rho=p_{\phi}\ $ with other components of the $T_{\mu\nu}$
vanishing. This can be interpreted as equal amount of oppositely
rotating photons along a circular path and the relations between
the parameters of the interior and exterior metrics
(\ref{interiorkmetric}-\ref{exteriorkmetric}) become $a=c',b=b'$
and $c=a'$. The shell has positive energy density for $a'>a$ and
the line singularity has again positive effective mass density. We
can find the solution where the shell with photons counter moving
along the $z$ direction with choosing $c=c'$ in (\ref{G33shell})
where the nonzero components of the $T_{\mu\nu}$ are $\rho=p_z$
and the relation between the parameters of the interior and
exterior metrics (\ref{interiorkmetric}-\ref{exteriorkmetric}) are
$a=b',b=a'$ and $c=c'$ . But this time, either shell or line
singularity has negative energy density \cite{ArikDelice}.

Now, let us discuss the two concentric thin shells around a line
singularity. The two shells seperate spacetime into three vacuum
regions which can be characterized by three different Levi-Civita
metrics in Kasner form (\ref{kasner}) with the metric parameters
$a,b,c,a',b',c',a'',b'',c'' \ $ which satisfy the relations
(\ref{abcrel}). These metrics can be combined as (\ref{metricABC})
where the metric is continuous but its first derivatives are
discontinuous with the functions given by
\begin{equation}\label{A2}
A(r)=\theta(r_1-r)(\frac{r}{r_1})^a+\theta(r-r_1)\theta(r_2-r)(\frac{r}{r_1})^{a'}+\theta(r-r_2)(\frac{r}{r_2})^{a''}(\frac{r_2}{r_1})^{a'},
\end{equation}
To obtain $B(r)$ replace $a$ with $b$ and to obtain $C(r)$ replace
$a$ with $c$ and multiply with $\alpha r_1$ in (\ref{A2}).
 The nonzero components of the Einstein tensor are:
\begin{equation}\label{G002}
G_{00}=(\frac{a'-a}{r_1})\ \delta(r-r_1)+(\frac{a''-a'}{r_2})\
\delta(r-r_2)
\end{equation}
\begin{equation}\label{Gzz2}
G_{zz}=(\frac{b-b'}{r_1})\ \delta(r-r_1)+(\frac{b'-b''}{r_2})\
\delta(r-r_2)
\end{equation}

\begin{equation}\label{G332}
G_{\phi\phi}=(\frac{c-c'}{r_1})\
\delta(r-r_1)+(\frac{c'-c''}{r_2})\ \delta(r-r_2).
\end{equation}
 Since the energy-momentum tensor satisfies  (\ref{traceless}), we have two infinitely thin photonic cylindrical shells
 with radii $r_1$ and $r_2$ where photons counter moving along a
 helical path around a line singularity. The first and second shells have positive energy density for $a'>a$
 and $a''>a'$. Thus if we have $a''>a'>a$ both of the two photonic shells have
 positive energy density.

 We can construct 3 photonic shells around a line singularity by
 choosing the continuous function A(r) as:
\begin{eqnarray}
 \label{A3}
 & &\ \ A(r)=\ \theta(r_1-r)(\frac{r}{r_1})^a+\theta(r-r_1)\ \theta(r_2-r)\ (\frac{r}{r_1})^{a'} \  \nonumber \\
& &
+\theta(r-r_2)\theta(r_3-r)(\frac{r}{r_2})^{a''}(\frac{r_2}{r_1})^{a'}+\theta(r-r_3)(\frac{r}{r_3})^{a'''}(\frac{r_3}{r_2})^{a''}(\frac{r_2}{r_1})^{a'}
.\end{eqnarray}

To obtain $B(r)$ replace $a$ with $b$ and to obtain $C(r)$ replace
$a$ with $c$ and multiply with $\alpha r_1$ in (\ref{A3}).
 The nonzero components of the Einstein tensor are calculated as:
\begin{equation}\label{G003}
 G_{00}=(\frac{a'-a}{r_1})\delta(r-r_1)+(\frac{a''-a'}{r_2})\delta(r-r_2)+(\frac{a'''-a''}{r_3})\delta(r-r_3),
\end{equation}
\begin{equation}\label{Gzz3}
 G_{zz}=(\frac{b-b'}{r_1})\delta(r-r_1)+(\frac{b'-b''}{r_2})\delta(r-r_2)+(\frac{b''-b'''}{r_3})\delta(r-r_3),
\end{equation}
\begin{equation}\label{G333}
 G_{00}=(\frac{c-c'}{r_1})\delta(r-r_1)+(\frac{c'-c''}{r_2})\delta(r-r_2)+(\frac{c''-c'''}{r_3})\delta(r-r_3).
\end{equation}
 For this case $T_{\mu \nu}$ again satisfy (\ref{traceless}) thus we
 have three photonic shells around a line singularity with photons moving along  helical path. If we have $a'''>a''>a'>a$ both of the three photonic shells have
 positive energy density.

 Thus, using this method, one can construct a solution where a line singularity  (for $a=b=0, c=1, \alpha < 1$ a cosmic string) at $r=0\ $ is surrounded by $n$ infinitely
 thin cylindrical photonic shells with photons  counter moving along a helical path. All shells satisfy the positive energy
 condition if  the metric parameters satisfy the condition
 \begin{equation}\label{a}
 a^{(n)}>\ldots>a'''>a''>a'>a.
\end{equation}
 Due to the relations (\ref{abcrel}) the maximum value that $a^{(n)}$ can have is
 1 and  in this case the exterior region of the multiple shells ($r>r_n$) is determined by the
 metric:
\begin{equation}\label{rindler}
    ds^2=-r^2dt^2+dr^2+dz^2+d\phi^2.
\end{equation}
This metric is the Rindler's metric \cite{rindler} which describes
static plane symmetric vacuum spacetime \cite{dasilva}. It
corresponds to a uniform gravitational field and test particles
are uniformly accelerated in this field whereas the Riemann tensor
is identically zero. Thus the line singularity may be surrounded
by multiple photonic shells until they all appear as an infinitely
long static plane wall.

We can also construct $n$ photonic shells with counter rotating
photons ($p_z=0$) or counter moving photons along axial direction
($p_\phi =0$) around a line singularity but if one shell has
positive energy density then the next one has negative energy
density  or vice versa thus these solutions cannot be physically
relevant. Notice that if interior singularity is a cosmic string
($a,b=0,c=1$), for a shell with counter rotating photons, the
exterior metric reduces to (\ref{rindler}) since for a counter
rotating shell $ a'=c,\ b'=b,\ c'=a.$

Our results can be summarized as follows. When a cosmic string is
surrounded by a single photonic shell composed of circular counter
rotating photons, from the outside the photonic shell looks like a
plane wall. However when the photons are helical, multiple
photonic shells are possible. There is no photonic shell  with
axial photons around a cosmic string. For a general line
singularity a single photonic shell is always possible. Multiple
photonic shells, however, require that the photons are helical,
i.e. multiple photonic shells are not possible if photons move
purely circularly or axially. In all cases the photonic shells may
be terminated by an outermost photonic shell, which, from the
outside looks like a planar wall.


\end{document}